# On the gene expression landscape of cancer

Augusto Gonzalez (1,2), Yasser Perera (3,4), Rolando Perez (1,5)

(1) University of Electronic Science and Technology of China, Chengdu, People Republic of China
(2) Institute of Cybernetics Mathematics and Physics, Havana, Cuba
(3) China-Cuba Biotechnology Joint Innovation Center, Yongzhou, People Republic of China
(4) Center of Genetic Engineering and Biotechnology, Havana, Cuba
(5) Center of Molecular Immunology, Havana, Cuba

**ABSTRACT**
A principal component analysis of the TCGA data for 15 cancer localizations unveils the following qualitative facts about tumors: 1) The state of a tissue in gene expression space may be described by a few variables. In particular, there is a single variable describing the progression from a normal tissue to a tumor. 2) Each cancer localization is characterized by a gene expression profile, in which genes have specific weights in the definition of the cancer state. There are no less than 2500 differentially-expressed genes, which lead to power-like tails in the expression distribution functions. 3) Tumors in different localizations share hundreds or even thousands of differentially expressed genes. There are 6 genes common to the 15 studied tumor localizations. 4) The tumor region is a kind of attractor. Tumors in advanced stages converge to this region independently of patient age or genetic variability. 5) There is a landscape of cancer in gene expression space with an approximate border separating normal tissues from tumors.

**INTRODUCTION**

Cancer is a very complex phenomenon [1]. From one side, the number of internal variables participating in relevant processes in a tissue is huge. The number of human genes, for example, is around 60000. Trying to infer the behavior of such a system from observations in a few hundred samples is very challenging. The general laws and state variables are not completely known or understood. From the other side, as in any aspect of life, ties with the environment are very strong.

Enormous coordinated efforts aimed at understanding basic aspects of cancer have led to many important results. A lot of information about genes, cells and tissues have been compiled and put into public databases [2-4]. The analysis of such information allowed the identification of mutation signatures, immune characteristics, etc [5-12].

From the theoretical point of view, current views to cancer include cell-intrinsic (i.e. genetic & epigenetic) and cell-extrinsic phenomena (i.e. micro-environment), as well as population genetics approaches with random drift and directional selection shaping, what has been coined as somatic evolution [13,14]. There is even a plausible hypothesis that cancer is an atavism, that is a cooperative state of multi-cellular organisms, prior to modern metazoans [15-19].

In spite of all this progress, a simple enough picture, quantifying the evolution of a normal tissue towards cancer is still lacking.

In this paper, we present a contribution to the qualitative perspective by stressing a few qualitative facts on cancer which come from the analysis of gene expression data. The title of the paper, "cancer gene expression landscape", indicates the aim at placing all tumors in a single plot and delineating the border between normal tissues and tumors.

The data, provided by The Cancer Genome Atlas (TCGA) project [4], is processed by standard Principal Component Analysis (PCA, [20-22]). In the paper, we avoid using models or elaborated theoretical constructions that could hide basic facts. Instead of this, we focus on general results following straightforwardly from the expression data.

**QUALITATIVE RESULTS**
We take tissue expression data for 15 cancer localizations from the TCGA project. This is a well curated database. Gene expressions are measured by sequencing the mRNA produced in the transcription process (RNA-seq, [23]). The data is in the number of fragments per kilo-base of gene length per mega-base of reads format (FPKM, [24]). We pick up localizations where there are at least 10 normal samples, and the number of tumor samples is greater than 160. The studied cases are shown in Table I.
(Insert Table I)

In the Methods section, the PCA used methodology is briefly explained. We take the mean geometric average over normal samples in order to define the reference expression for each gene, and normalize accordingly to obtain the differential expressions, $\bar{e} = e/e_{ref}$. Finally, we take the base 2 logarithm, $\hat{e} = Log2 (\bar{e})$, to define the fold variation. Besides reducing the variance, the logarithm allows treating over- and sub-expression in a symmetrical way. The covariance matrix is defined in terms of $\hat{e}$. We forced the reference for the PC analysis to be at the center of the cloud of normal samples. This is what actually happens in a population, where most individuals are healthy and cancer situations are rare. The eigenvectors of the covariance matrix define the PC axes: PC1, PC2, etc, and projection over them define the new state variables. By definition, PC1 captures the highest fraction of the total variance in the sample set.

**Statement 1. The state of a tissue in gene expression space may be described by a few variables. In particular, there is a single variable describing the progression from a normal tissue to a tumor.**
Fig. 1 shows the PCA results for three of the studied tissues: Kidney renal papillary cell carcinoma (KIRP), Lung squamous cell carcinoma (LUSC) and Liver hepatocellular carcinoma (LIHC). In the left panel, the distribution of normal and tumor samples in the (PC1, PC2) plane suggests that there are well defined normal and tumor regions, and that the first PC1 variable may discriminate between a normal sample and a tumor. PC1 can thus be labeled as the cancer axis.
(Insert Fig. 1)

The projection along PC1 may be used to quantify the progression from a normal tissue to a tumor. Comparison with other qualifications like the tumor stage is interesting. We shall come back to this point below.

The right panel, on the other hand, evaluates the fraction of total variance captured by the first n PC variables. In LUSC, for example, the first three components account for 74% of the variance. Notice that with a few such variables we may describe to a considerable extent the complexity of LUSC.

This reduced number of variables, well below the number of constituent genes, may be taken as the effective number of degrees of freedom of the complex system represented by a tissue.

The fraction of variance depends on the sample set, and may vary if we enlarge or reduce the set. Thus, we should take the results as approximate, semi quantitative ones, that shall improve as the sample size is enlarged.

However, in spite of the statistical origin of the new variables we may use them to describe the actual state of a given sample. Each unitary vector along any of the PC axes defines an expression profile, with a meaning. The fact that with 6 - 10 such variables we may account for 80 % or more of the data dispersion means that the effective dimensionality of the state space is much less than it seems. Genes do not take arbitrary expressions, but act in a concerted way. Groups of genes, metagenes or sub-networks express themselves as profiles.

**Statement 2. Each cancer localization is characterized by a gene expression profile, in which genes have specific weights in the definition of the cancer state.**

Let us call $\mathbf{v_1}$ the eigenvector along PC1 (boldface denotes vectors). We showed above that the PC1 axis accounts for the largest fraction of variance and that projection along it may be taken as an indicator of the malignant state. For a given sample with fold expression vector $\hat{\mathbf{e}}$, the projection $x_1$ over the PC1 axis is precisely defined as:

$$x_1 = \hat{\mathbf{e}} \cdot \mathbf{v_1} = \sum \hat{e}_i \, v_{1i}.$$

The $\mathbf{v_1}$ vector may be thought to provide a metagene [25] or gene expression profile of cancer in the tissue, i.e. the set of over- or under-expressed genes (and their relative importance) that define the cancer state. $v_{1i}$ is the weight of gene i in the definition of the cancer state. A positive sign means that the gene is over-expressed in the tumor.

Fig. 2 left panel shows the 30 genes with the largest contributions to $\mathbf{v_1}$ in the same tumors represented in Fig. 1. Note the signs, positive and negative indicating over- and under-expression respectively. The components of the $\mathbf{v_1}$ vector define the weights of these genes in the definition of the cancer state. In principle, because of their large weights, these 30 genes could be used as cancer biomarkers, however their specific roles in each tissue deserve further study. In LUSC, for example, the gene with the largest weight is SFTPC, a silenced gene with an important role in lung homeostasis [26,27]. The analogous genes in KIRP and LIHC are Uromodulin (UMOD) and Cytochrome P450 family 1 subfamily A member 2 (CYP1A2), respectively. This analysis is promising and, to the best of our knowledge, have not been sufficiently exploited so far.
(Insert Fig. 2)

The central and right panels of Fig, 2 show genes, ordered according to their differential expressions, for the centers of the tumor clouds (geometric averages over tumor samples). Only the tails with higher over- or under-expressions are shown. Notice that these tails contains a few thousands of genes, the rest of the 60000 genes are not differentially expressed. The distributions are not symmetrical. Whereas Kidney Papillary Cell Carcinoma (KIRP) is dominated by silenced genes, LUSC has nearly equal proportions of under- and over-expressed genes, and in Liver Hepatic Cell Carcinoma (LIHC) the over-expressed genes are more numerous. Log-log plots stress that the tails exhibit a power-like (Pareto, [28,29]) behavior, i.e. the number of genes with differential expression greater than a given value is an inverse power of the expression.

**Statement 3. Tumors in different localizations share hundreds or even thousands of differentially expressed genes. There are 6 genes common to 15 tumor localizations.**

For each localization, we select the most significant 2500 genes with the largest contributions to the vector $\mathbf{v_1}$ along the PC1 direction defining the cancer state. This number, although arbitrary, is dictated by the previous results on the expression distribution function. Let us stress that these are genes with significant differential expressions and great importance in the definition of the cancer state.

Table II shows the number of shared genes for pairs of localizations. Notice that these numbers vary in the interval between 314 and 1889. Large numbers of shared genes are characteristic of tumors in the same organ but originating in different cells (lung, kidney). However, there are also tumors sharing unexpectedly large numbers of genes. For example, tumors in the uterine corpus (UCEC) and bladder (BLCA) share more than 1300 genes.
(Insert Table II)

Let us stress that there are 49 genes common to a group of 11 tumors, PRAD-LUSC-LUAD-UCEC-BLCA-ESCA-BRCA-HNSC-COAD-READ-STAD, and six genes common to all of the studied localizations. They are MMP11 (+), C7 (-), ANGPTL1 (-), UBE2C (+), IQGAP3 (+) and ADH1B (-). Their differential expressions are very similar in all of the studied tumors. The signs added in parenthesis mean that the gene is over- or under-expressed in tumors.

The six identified pan-cancer genes have been recently pointed out as playing a significant role in many cancers [30-35]. It is noteworthy that these genes are straightforwardly related to cancer hallmarks [36,37]: i.e. invasion, suppression of the immune response, angiogenesis, proliferation and changes in metabolism.

Shared genes among groups of tumors open the question about universal therapies for these groups.

Below, we notice that pan cancer genes (the six ones common to all of the tumors) play a role in both tissue differentiation and in the definition of the border between normal tissues and tumors. In addition, the number of shared genes seems to be related to the proximity of tumors in the expression space.

**Statement 4. The tumor region is a kind of attractor. Tumors in advanced stages converge to this region independently of patient age or genetic variability.**
As may be seen in Fig. 1, regions corresponding to normal and tumor samples are well defined and partially disjoint in the expression space. The sample variability comes from genetic differences, patient ages and the evolution history of each individual. The fact that, in spite of variability, regions are well defined in expression space is in favor of the attractor paradigm of cancer [38-40] in which the cancer region should be the region of confluence of all somatic evolution trajectories out of the normal area. In a very reductionist view one may think, for example, about normal functioning and cancer as two stable solutions of a global gene regulatory network [41,42].

We may conduct a more precise test of the attractor hypothesis by studying the dependence of the distribution functions on patient age. Let us consider, again, LUSC as an example. According to age, in LUSC we may define 4 subgroups of samples: Normal Young (NY), Normal Old (NO), Tumor Young (TY) and Tumor Old (TO). These subgroups are in some sense arbitrarily defined. First, the label "normal" refers to a pathologically normal sample from a patient with a tumor. Second, in the example "young" labels samples where age is lower than 62 years. These conditions are dictated by the availability of samples, as in any statistical analysis.

Nevertheless, the results are very interesting. The (over-) expression distribution function is visualized in Fig. 3. We compute (mean geometric) averages over the NY, NO, TY and TO subgroups, and use the NY values as references in order to define normalized (differential) expressions in the remaining subgroups: $\bar{e}_{NO}$, $\bar{e}_{TY}$, and $\bar{e}_{TO}$. These vectors characterize the centers of their respective clouds of samples. Genes are sorted with regard to their normalized expression values.
(Insert Fig. 3)

There are deviations for a reduced number of genes in the NO group with regard to the NY reference. This may be taken as a consequence of aging. In tumors, however, deviations are much larger. There are around 1000 over-expressed genes with $|\bar{e}| > 5$.

Most striking is the similarity between the distribution functions in the TY and TO subgroups. That is, for tumors the distribution function in the final state is nearly independent of the age when tumor initiates. This is an argument in favor of the attractor hypothesis. Similar results (not shown) are obtained for the sub-expression tail.

A slightly different test comes from considering a second "time" or progression variable: the clinically determined tumor stage. It is a qualification given to the tumor at the moment of diagnosis, but in some sense it quantifies also the somatic evolution once the portion of the tissue acquires the tumor condition. Fig. 4 shows the distribution of tumors by stages in Clear Cell Kidney Carcinoma (KIRC). Normal tissues are represented by blue points, whereas tumors are drawn in red. The four panels refer to the four stages: I, II, III and IV. Blue points are in the four panels, but only red points with the corresponding stage are included in each panel.
(Insert Fig. 4)

We use a contour plot and colors to visualize the total density of points in the state space. Two regions of maximal density are apparent, corresponding to normal states and an optimal region for tumors. Naively, one expects that tumors move along the transition region from the normal to the tumor region as the stage evolves from I to IV. In the actual measurements, we don't track individual tumors as function of stages, but get pictures of different tumors at different stages. Thus, in the initial stages we should observe a fraction of red points captured in the transition region, whereas in the final stages, most tumors should be concentrated in the optimal region. This is what actually follows from the figure, again supporting the attractor paradigm. We may speculate that the optimal region could be related to a region of maximal fitness for the tumor in the given tissue.

The intuition induces us to relate the tumor stage to the coordinate along the tumor axis PC1. The correspondence, however, is not exact. Although there seems to be a correlation between stage and mean displacement towards the tumor region, many tumors in the initial stages are already at the center of the cloud. This could be related to the fact that the observed distribution of samples is probably related to the fitness distribution and the transition region should be a low-fitness zone.

**Statement 5. There is a landscape of cancer in gene expression space with an approximate border separating normal tissues from tumors.**
We consider the central goal of the paper: i.e. to draw a picture in which both normal tissues and tumors in different localizations are represented. In a way, this is a picture involving tissue differentiation and cancer. It is not surprising that pan cancer genes will play a role in both processes.

We shall use the $\mathbf{e}_{normal}$ and $\mathbf{e}_{tumor}$ (mean geometric) averages for each localization in order to define cloud centers. The reference is to be computed by averaging over all normal expression vectors. Then, the ê magnitudes and the covariance matrix are obtained, and the latter diagonalized.

The first aspect to be stressed is that the first two PCs accounts only for 37 % of the total data variance. The relative importance of these two variables is not so apparent as in the case of individual tissues. This is probably due to the big dispersion of the data for normal tissues, related to tissue differentiation, sometimes even larger than separations between a normal tissue and the respective tumor.

As a consequence of the dispersion of normal tissues, we do not have a "cancer axis" or direction,

as in individual tissues. In order to draw a frontier between normal and tumor regions, we shall include higher PCs. The next component, PC3 accounts for 12 % of the data variance.

We show in Fig. 5 the (PC1, PC3) plane, which indeed suggests that there is a border. Actually, the regions and the border are high dimensional, but the 2D figure captures the essential features. We may baptize this figure as the "approximate normal vs cancer" or "tissue differentiation vs cancer" plane. It is apparent from the figure, that the transition from a normal tissue to the corresponding tumor implies crossing the border, and involves simultaneous displacements along the PC1 and PC3 axes.
(Insert Fig. 5)

The unitary vectors along these axes allow the identification of genes with the highest weights. It is very interesting that pan cancer genes are among the most important genes in these vectors. For example, ADH1B and UBE2C are included in the set of 8 most important genes along PC1: PI3 (+), ADH1B (-), MYBL2 (+), UBE2C (+), ALB (-), CEACAM5 (+), CST1 (+) and MMP1 (+).

A more detailed analysis of the border between normal tissues and tumors is required. In the present paper, we limit ourselves to draw the global picture, and leave this analysis for a future work.

We would like to notice that distances between pairs of tumors in gene expression space are inversely correlated with the number of shared genes which define the cancer state. This fact is partially reflected in Fig. 5 because the distances in this figure are not true distances, but projections.

**DISCUSSION**
We performed PC analysis of gene expression data for 15 tumors. Our results are approximate and semi-quantitative, in the sense that they could be modified if a larger data set become available, but at the same time they are simple, general and unbiased, in the sense that no modeling or elaborated mathematical treatments are used. We try to keep interpretation of results as close as possible to the facts.

Both somatic evolution in a normal tissue, and the transition to a tumor state involves the modification of thousands of genes. However, these genes do not act independently, but in a concerted way. The number of relevant PC coordinates for a tissue (or a portion of it), which seems to be around 10, may be interpreted as the effective number of degrees of freedom of the biological system.

These variables, although of statistical origin, can be used to describe the state and evolution of the tissue, in particular one of the variables measures the progression from normal to cancer state (the projection on PC1, the cancer axis) and allows the definition of a profile of genes involved in this progression.

The data seem to support the theory of cancer as an attractor, that is once a portion of the tissue escapes from the normal region it is driven to the cancer basin of attraction.

The existence of a ranking of genes in the definition of the cancer state, allows us to speculate about the possibility to stop progression if the tumor is diagnosed in the initial stage. Indeed, what would happen if we target a few of the most significant genes in the cancer profile? Could this kind of intervention induce a rearrangement of the whole profile preventing the tumor to evolve to more advanced stages? This is an interesting question with possible implications for therapy.

Questions are also raised in relation to the second important conclusion of our paper, related to the overall landscape in gene expression space: not only individual tissues are separated from their respective tumors, but the set of normal tissues are separated from the set of tumors. In a very rough analysis, we noticed that pan cancer genes (i.e, common to all 15 tumors) are involved both in tissue differentiation and in the definition of the border. One possible, very interesting question is the following: what could this relationship tell us about tumors in early childhood, which initiate in the developmental period? On the other hand, one can imagine therapies that make use of the complex landscape represented in Fig. 5. For example, could we target the main genes connecting a tumor and a different nearby (in GE space) normal tissue? Would this intervention reduce the tumor fitness in the original tissue leading to a regression?

In the paper, we focused on qualitative aspects. However, there is the possibility to quantify and to model some aspects of cancer. For example, notice that from figures like Fig. 1 we can estimate the dimensions of the normal and cancer regions in gene expression space and the distance between their centers. From this data, and the statement that progression is described by a single variable one may possibly devise a one dimensional model for tumorigenesis in a given tissue [43].

Work along some of these directions is in progress [44,45].

**METHODS**

The results of the paper are based on the analysis of TCGA data for gene expression in FPKM format. The number of genes is 60483. This is the dimension of matrices in the Principal Component analysis.

We selected the 15 cancer types shown in Table I on the basis of two conditions: i) The number of normal samples is greater than 10, and ii) The number of tumor samples is greater than 160.

We show in Fig. S1 expressions from a typical data file (PRAD case). Notice that there are around 28000 not transcribed genes (expression exactly zero), and only around 25000 genes with expression above 0.1.

Usually, in order to compute the average expression of a gene the median or the geometric mean are used. We prefer geometric averages, but then the data should be slightly distorted to avoid zeroes. To this end, we added a constant 0.1 to the data. By applying this regularization procedure, genes identified as relevant could be under question if the differential expression is relatively low and their expression in normal tissues is near zero. As we are mainly interested in the strongly over- or under-expressed genes, they are out of the question.

For each cancer localization, we take the mean geometric average over normal samples in order to define the reference expression for each gene, $e_{ref}$. Then the normalized or differential expression is defined as: $\bar{e} = e/e_{ref}$. The fold variation is defined in terms of the base 2 logarithm $\hat{e} = \log_2(\bar{e})$. Besides reducing the variance, the logarithm allows treating over- and sub-expression in a symmetrical way.

Deviations and variances are measured with respect to $\hat{e} = 0$. That is, with respect to the average over normal samples. This election is quite natural, because normal samples are the majority in a population, individuals with cancer are rare.

With these assumptions, the covariance matrix is written:

$$\sigma^2_{ij} = \sum \hat{e}_i(s)\, \hat{e}_j(s) / (N_{samples}-1)$$

where the sum runs over the samples, s, and $N_{samples}$ is the total number of samples. $\hat{e}_i(s)$ is the fold variation of gene i in sample s.

As mentioned, the dimension of matrix $\sigma^2$ is 60483. By diagonalizing it, we get the axes of maximal variance: the Principal Components (PCs). They are sorted in descending order of their contribution to the variance.

In LUSC, for example, PC1 accounts for 67% of the variance. This large number is partly due to our choice of the reference, $\hat{e} = 0$, and the fact that most of the samples are tumors. The reward is that PC1 may be defined as the cancer axis. The projection over PC1 defines whether a sample is classified as normal or tumor.

The next PCs account for a smaller fraction of the variance. PC2 is responsible of 4%, PC3 of 3%, etc. Around 10 PCs are enough for an approximate description of the region of the gene expression space occupied by the set of samples.

Thus, we need only a small number of the eigenvalues and eigenvectors of $\sigma^2$. To this end, we use a Lanczos routine in Python language, and run it in a node with 2 processors, 12 cores and 64 GB of RAM memory. As a result, we get the first 100 eigenvalues and their corresponding eigenvectors.

**Conflict of interes**
The authors have no conflict of interests.

**Acknowledgments**
A.G. acknowledges the Cuban Program for Basic Sciences, the Office of External Activities of the Abdus Salam Centre for Theoretical Physics, and the University of Electronic Science and Technology of China for support. The research is carried on under a project of the Platform for Bio-informatics of BioCubaFarma, Cuba. The data for the present analysis come from the TCGA Research Network: https://www.cancer.gov/tcga.

**Author contributions**
Conceptualization & Investigation, all authors; Software and formal analysis, A.G.; Writing – Original Draft, A.G.; Writing – Review & Editing, all authors.

# REFERENCES

[1] Hiroaki Kitano (2002). Computational systems biology. Nature 420: 206-210.

[2] GeneCards, The Human Gene Database: https://www.genecards.org.

[3] The Human Protein Atlas: https://www.proteinatlas.org.

[4] The TCGA Research Network: https://www.cancer.gov/tcga.

[5] Matthew H. Bailey, Collin Tokheim, Eduard Porta-Pardo, et al (2018). Comprehensive Characterization of Cancer Driver Genes and Mutations. Cell 173, 371–385.

[6] Kuan-lin Huang, R. Jay Mashl, Yige Wu, et al (2018). Pathogenic Germline Variants in 10,389 Adult Cancers. Cell 173, 355–370.

[7] Vésteinn Thorsson, David L. Gibbs, Scott D. Brown, et al (2018). The Immune Landscape of Cancer. Immunity 48, 812–830.

[8] Li Ding, Matthew H. Bailey, Eduard Porta-Pardo, et al (2018). Perspective on Oncogenic Processes at the End of the Beginning of Cancer Genomics. Cell 173, 305–320.

[9] The ICGC/TCGA Pan-Cancer Analysis of Whole Genomes Consortium (2020). Pan-cancer analysis of whole genomes. Nature 578: 82-93.

[10] Moritz Gerstung, Clemency Jolly, Ignaty Leshchiner, et al (2020). The evolutionary history of 2,658 cancers. Nature 578: 122-128.

[11] PCAWG Transcriptome Core Group, Claudia Calabrese, Natalie R. Davidson, et al (2020). Genomic basis for RNA alterations in cancer. Nature 578: 129-136.

[12] Matthew A. Reyna, David Haan, Marta Paczkowska, et al (2020). Pathway and network analysis of more than 2500 whole cancer genomes. Nature Communications 11: 729.

[13] P.C. Nowell (1976). The clonal evolution of tumor cell populations. Science 194: 23-28.

[14] Martin A. Nowak, Franziska Michor, and Yoh Iwasa (2003). The linear process of somatic evolution. PNAS 100: 14966 –14969.

[15] P.C.W. Davies and C. H. Lineweaver (2011). Cancer tumors as Metazoa 1.0: tapping genes of ancient ancestors. Phys. Biol. 8(1): 015001.

[16] Tomislav Domazet-Lošo and Diethard Tautz (2010). Phylostratigraphic tracking of cancer genes suggests a link to the emergence of multicellularity in metazoa. BMC Biology 8: 66.

[17] Charles H. Lineweaver, Paul C. W. Davies and Mark D. Vincent (2014). Targeting cancer's weaknesses (not its strengths): Therapeutic strategies suggested by the atavistic model. Bioessays 36: 827–835.

[18] Luis Cisneros, Kimberly J. Bussey, Adam J. Orr, et al (2017). Ancient genes establish stress-induced mutation as a hallmark of cancer. PLoS ONE 12(4): e0176258.

[19] Anna S Trigos, Richard B Pearson, Anthony T Papenfuss, and David L Goode (2019). Somatic mutations in early metazoan genes disrupt regulatory links between unicellular and multicellular genes in cancer. ELife 8: e40947.

[20] Svante Wold, Kim Esbensen, and Paul Geladi (1987). Principal component analysis. *Chemometrics and intelligent laboratory systems* 2(1-3): 37-52.

[21] Jake Lever, Martin Krzywinski & Naomi Altman (2017). Principal component analysis. NATURE METHODS 14: 641-642.

[22] Markus Ringnér (2008). What is principal component analysis?. NATURE BIOTECHNOLOGY 26: 303-304.

[23] Zhong Wang, Mark Gerstein, and Michael Snyder (2009). RNA-Seq: a revolutionary tool for transcriptomics. Nat. Rev. Genet. 10(1): 57–63.

[24] C. Trapnell et al. (2010). Transcript assembly and quantification by RNA-Seq reveals unannotated transcripts and isoform switching during cell differentiation. Nature Biotechnology 28: 511-515.

[25] Erich Huang, Skye H Cheng, Holly Dressman, et al (2003). Gene expression predictors of breast cancer outcomes. The Lancet 361(9369): 1590-1596.

[26] Jeffrey A. Whitsett, Timothy E. Weaver (2002). Hydrophobic surfactant proteins in lung function and disease. N. Engl. J. Med. 347: 2141-2148.

[27] Surafel Mulugeta, Michael F. Beers (2006). Surfactant protein C: Its unique properties and emerging immunomodulatory role in the lung. Microbes and Infection 8: 2317-2323.

[28] M.E.J. Newman (2005). Power laws, Pareto distributions and Zipf's law. Contemporary Physics 46: 323-351.

[29] V. A. Kuznetsov,1 G. D. Knott and R. F. Bonner (2002). General Statistics of Stochastic Process of Gene Expression in Eukaryotic Cells. Genetics 161: 1321–1332.

[30] Qi-Gang Li, Yong-Han He, Huan Wu, et al (2017). A Normalization-Free and Nonparametric Method Sharpens Large-Scale Transcriptome Analysis and Reveals Common Gene Alteration Patterns in Cancers. Theranostics 7(11): 2888-2899.

[31] Carmine Carbone, Geny Piro, Valeria Merz et al (2018). Angiopoietin-Like Proteins in Angiogenesis, Inflammation and Cancer. Int. J. Mol. Sci. 19: 431.

[32] Vahid Afshar-Kharghan (2017). The role of the complement system in cancer. J. Clin. Invest. 127(3): 780–789.

[33] Ying Yang, Wei Zhao, Qing-Wen Xu et al (2014). IQGAP3 Promotes EGFR-ERK Signaling and the Growth and Metastasis of Lung Cancer Cells. PLoS ONE 9(5): e97578.

[34] Emily Gobin, Kayla Bagwell, John Wagner et al (2019). A pan-cancer perspective of matrix metalloproteases (MMP) gene expression profile and their diagnostic/prognostic potential.

[35] Hassan Dastsooz, Matteo Cereda, Daniela Donna and Salvatore Oliviero (2019). A

Comprehensive Bioinformatics Analysis of UBE2C in Cancers. Int. J. Mol. Sci. 20: 2228.

[36] Douglas Hanahan and Robert A. Weinberg (2000). The Hallmarks of Cancer. Cell 100: 57–70.

[37] Douglas Hanahan and Robert A. Weinberg (2011). Hallmarks of Cancer: The Next Generation. Cell 144: 646-674.

[38] S.A. Kauffman (1969). Metabolic stability and epigenesis in randomly constructed genetic nets. Journal of Theoretical Biology 22(3): 437-467.

[39] S. Huang, G. Eichler, Y. Bar-Yam and D.E. Ingber (2005). Cell fates as high-dimensional attractor states of a complex gene regulatory network. Phys Rev Lett. 94(12):128701.

[40] Sui Huang, Ingemar Ernberg, and Stuart Kauffman (2009). Cancer attractors: A systems view of tumors from a gene network dynamics and developmental perspective. Semin. Cell. Dev. Biol. 20(7): 869–876.

[41] Guy Karlebach and Ron Shamir (2008). Modelling and analysis of gene regulatory networks. NATURE REVIEWS 9: 771-780.

[42] Frank Emmert-Streib, Matthias Dehmer and Benjamin Haibe-Kains (2014). Gene regulatory networks and their applications: understanding biological and medical problems in terms of networks. Frontiers in Cell and Developmental Biology 2: 38.

[43] Roberto Herrero, Dario Leon and Augusto Gonzalez (2020). Levy model of cancer. arXiv:1507.08232 (unpublished).

[44] Frank Quintela and Augusto Gonzalez (2020). Estimating the number of available states for normal and tumor tissues in gene expression space. arXiv:2005.02271 (unpublished).

[45] Augusto Gonzalez, Joan Nieves, Maria Luisa Bringas and Pedro Valdes Sosa (2020). Gene expression rearrangements denoting changes in the biological state. arXiv:1706.09813 (unpublished).

| Cancer type (TCGA notation) | Normal samples | Tumor samples |
| --- | --- | --- |
| KIRP | 32 | 289 |
| KIRC | 72 | 539 |
| PRAD | 52 | 499 |
| LUSC | 49 | 502 |
| LUAD | 59 | 535 |
| UCEC | 23 | 552 |
| BLCA | 19 | 414 |
| COAD | 41 | 473 |
| ESCA | 11 | 160 |
| LIHC | 50 | 374 |
| STAD | 32 | 375 |
| BRCA | 112 | 1096 |
| HNSC | 44 | 502 |
| THCA | 58 | 510 |
| READ | 10 | 167 |

**Table I. The set of data analyzed in the paper**.

|      | KIRC | PRAD | LUSC | LUAD | UCEC | BLCA | COAD | READ | ESCA | STAD | BRCA | HNSC | LIHC | THCA |
|------|------|------|------|------|------|------|------|------|------|------|------|------|------|------|
| KIRP | 1213 | 534  | 642  | 625  | 696  | 643  | 543  | 545  | 532  | 464  | 646  | 523  | 459  | 600  |
| KIRC |      | 488  | 507  | 551  | 493  | 472  | 537  | 516  | 558  | 455  | 468  | 576  | 462  | 475  |
| PRAD |      |      | 561  | 608  | 774  | 789  | 621  | 712  | 573  | 597  | 730  | 581  | 485  | 481  |
| LUSC |      |      |      | 1454 | 857  | 932  | 634  | 591  | 1011 | 699  | 871  | 805  | 539  | 531  |
| LUAD |      |      |      |      | 863  | 883  | 801  | 731  | 778  | 803  | 908  | 725  | 628  | 528  |
| UCEC |      |      |      |      |      | 1313 | 710  | 831  | 702  | 784  | 1118 | 691  | 502  | 583  |
| BLCA |      |      |      |      |      |      | 813  | 935  | 817  | 965  | 1101 | 785  | 615  | 562  |
| COAD |      |      |      |      |      |      |      | 1889 | 682  | 994  | 722  | 622  | 512  | 517  |
| READ |      |      |      |      |      |      |      |      | 587  | 951  | 761  | 567  | 450  | 518  |
| ESCA |      |      |      |      |      |      |      |      |      | 977  | 697  | 1069 | 639  | 502  |
| STAD |      |      |      |      |      |      |      |      |      |      | 700  | 859  | 630  | 433  |
| BRCA |      |      |      |      |      |      |      |      |      |      |      | 707  | 555  | 589  |
| HNSC |      |      |      |      |      |      |      |      |      |      |      |      | 612  | 485  |
| LIHC |      |      |      |      |      |      |      |      |      |      |      |      |      | 314  |

**Table II. Number of common differentially-expressed genes in pairs of localizations**.

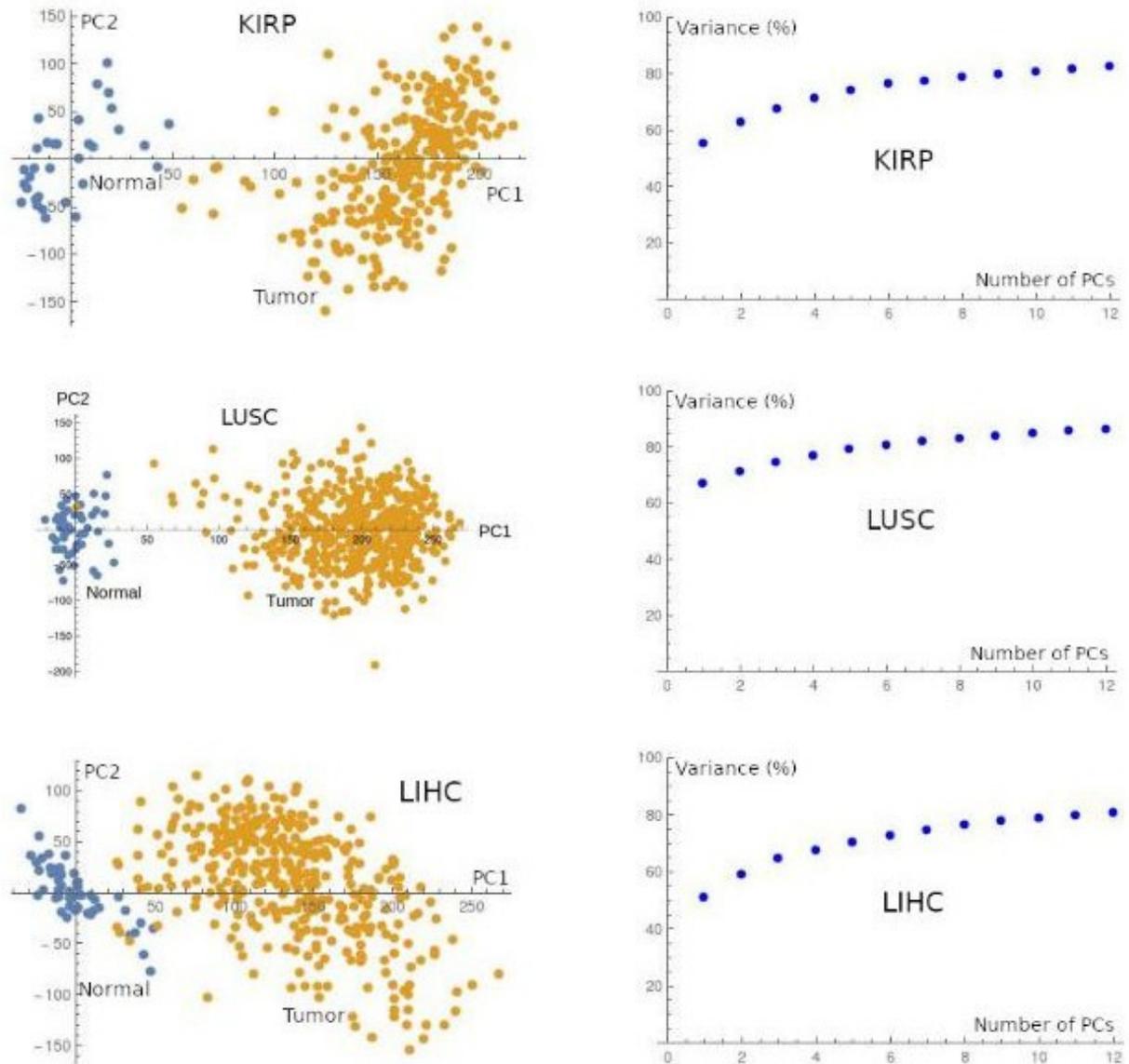

**Fig. 1. Principal Component Analysis of the TCGA gene expression data for three of the studied tumors**. The left panel contains the (PC1, PC2) plane, whereas the right panel shows the variance captured by the first n PCs.

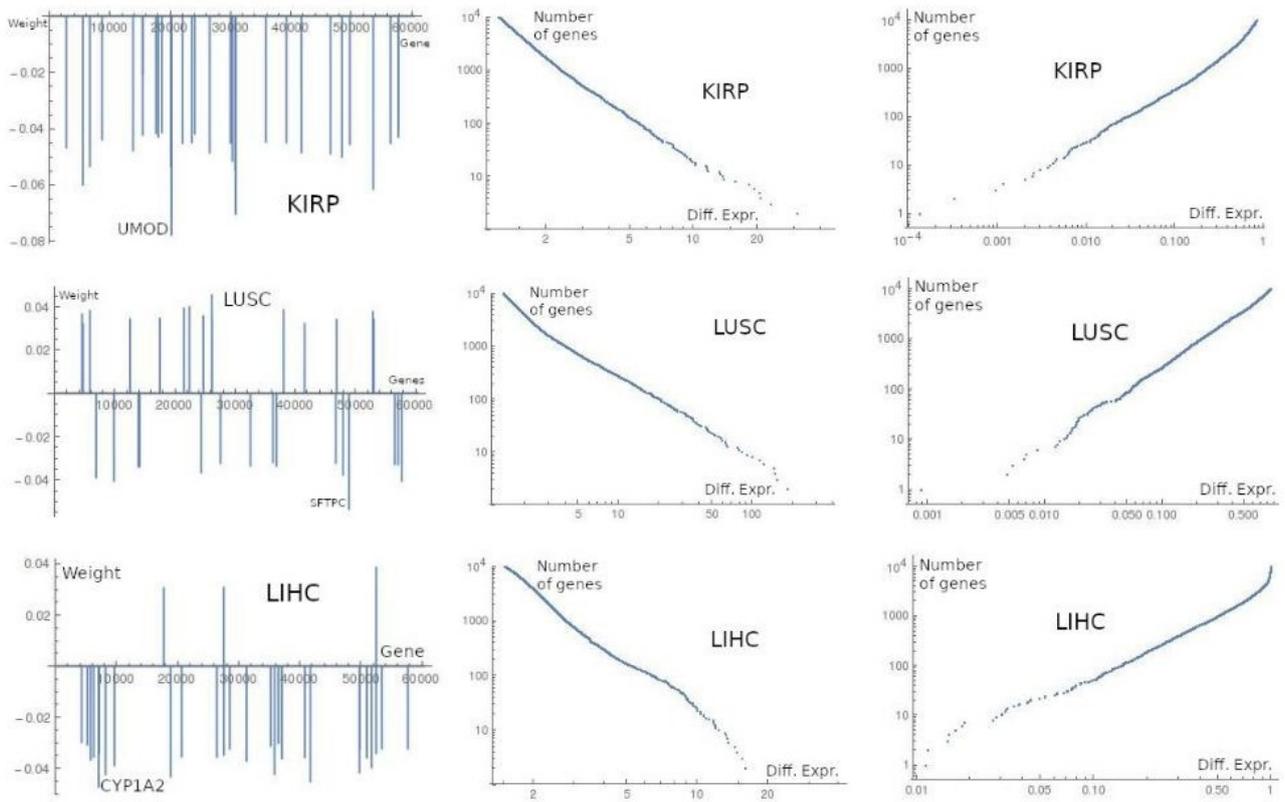

**Fig. 2. Left panel**. The 30 genes with highest weights in the definition of the cancer state. The same tumors as in Fig. 1 are used as examples. The numbering of genes is the one used in the TCGA data. To simplify drawing, the contributions of the remaining genes are set to zero. Positive signs correspond to over-expressed, and negative to sub-expressed genes. **Central and right panels**. Integrated gene expression distribution functions, over- and under-expression tails are shown.

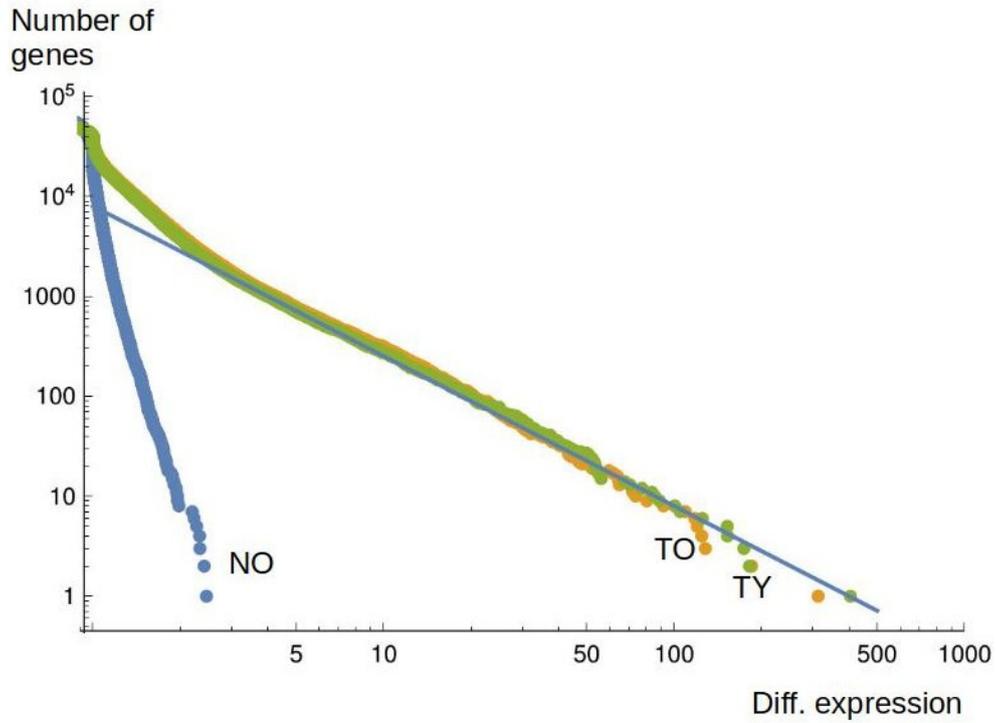

**Fig. 3 Integrated gene (over) expression distribution functions in LUSC**. According to age, samples are grouped into four sets: Normal Young (NY), Normal Old (NO), Tumor Young (TY) and Tumor Old (TO). The average over the NY set is used to define reference values to normalize the expressions. Each set of points represents the average over the respective group.

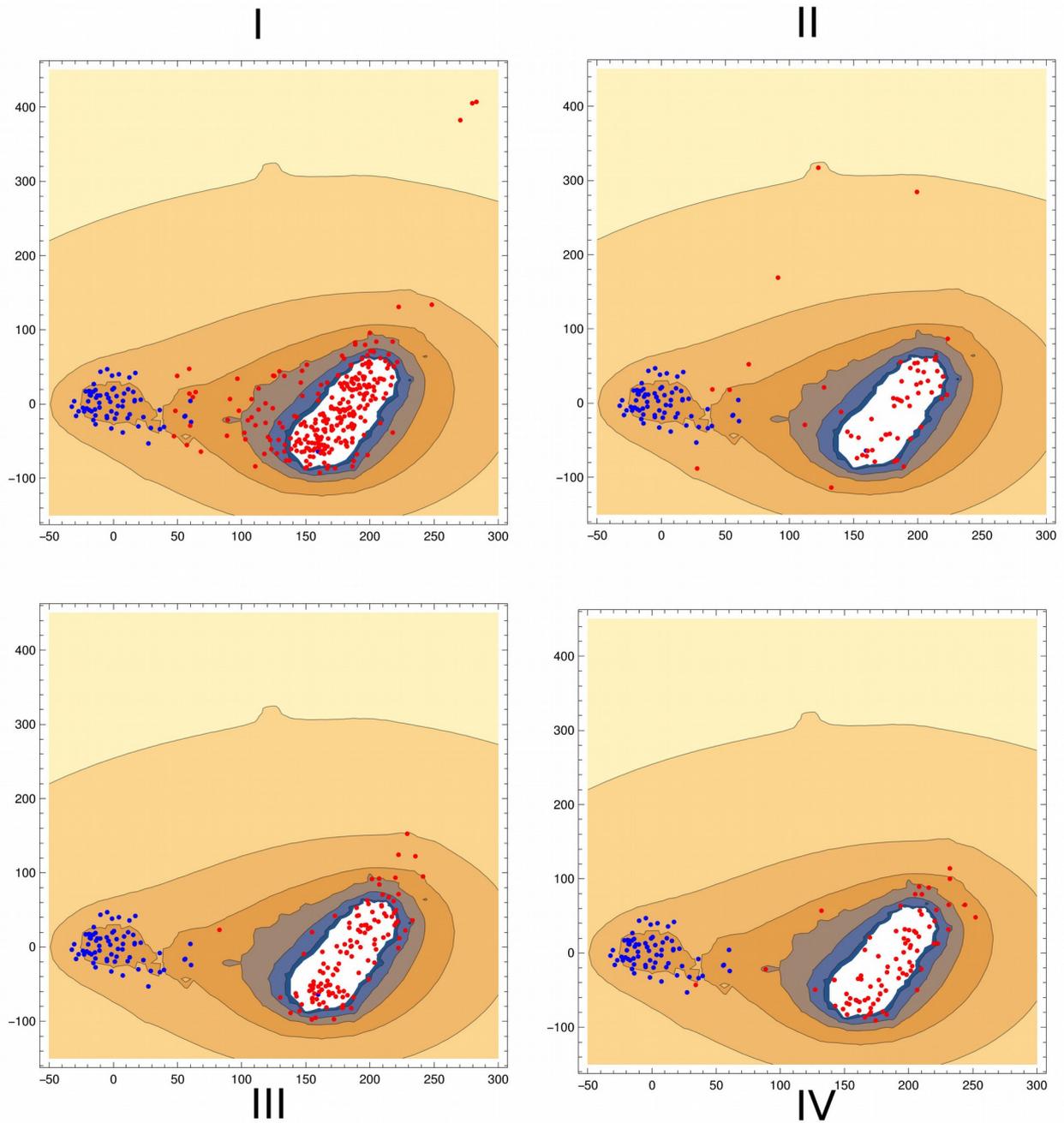

**Fig. 4. Stages in the evolution of tumors in Clear Cell Kidney Cancer (KIRC)**. Blue points are normal tissues (included in the four panels), whereas red points are tumors in a given stage of evolution. Contours represent the total density of points. Stage I seems to be "transitional", there are many points traveling along the intermediate region. On the other hand, stages II, III and IV are "final", in the sense that most of the tumors are concentrated in the high density region. This picture reinforces the attractor paradigm of cancer. We may speculate that the attractor is the region of the state space with maximal fitness for the tumor in the given tissue.

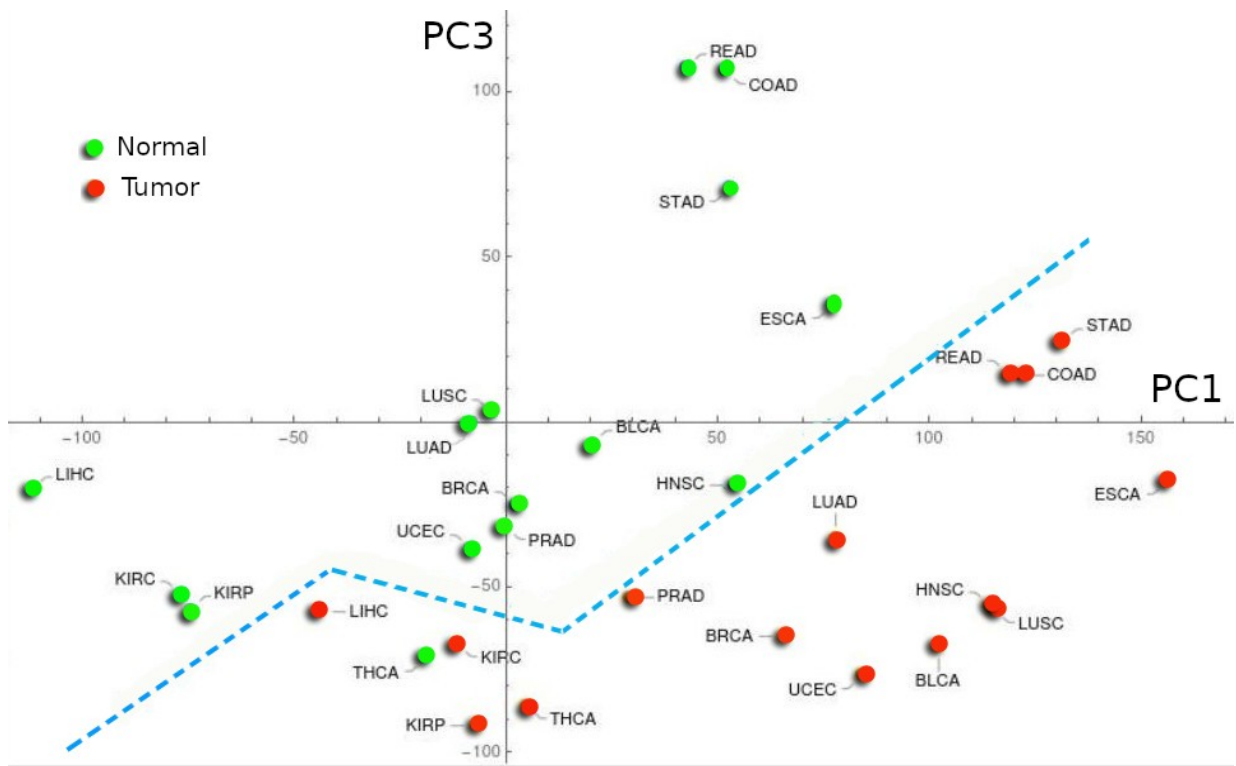

**Fig. 5. The gene expression landscape in the (PC1, PC3) plane**. Each point in the diagram represents the average of samples in a given localization. For simplicity, normal tissues are labeled with the corresponding tumor indexes. The approximate border between the normal and tumor regions is drawn.

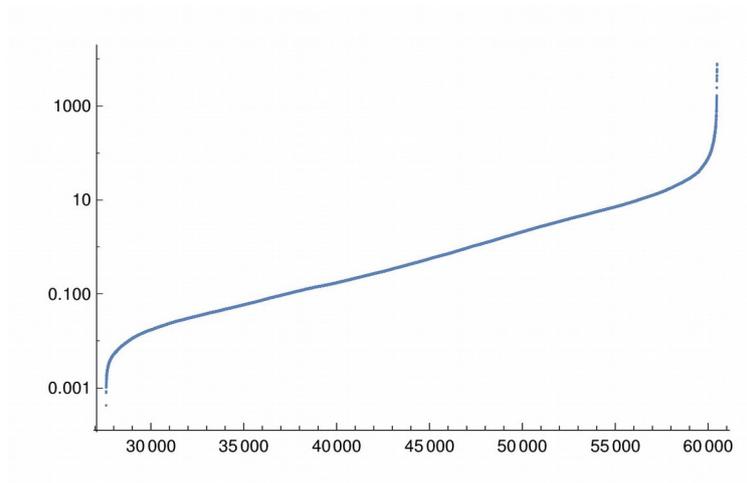

**Fig. S1 (Supplementary Fig. 1). Range of values in a typical data file**. Roughly half of the 60000 genes are not transcribed.